\documentclass[pra,aps]{revtex4}
\usepackage{graphicx}
\usepackage{colordvi}
\begin{document}
\title{Averaged dynamics of soliton molecules in
dispersion-managed optical fibers}

\author{S. M. Alamoudi$^1$, U. Al Khawaja$^2$, and B. B. Baizakov$^3$}
\affiliation{$^1$ Department of Physics, King Fahd University of
Petroleum and Minerals, Dhahran 31261, Saudi Arabia \\
$^2$ Physics Department, United Arab Emirates
University, P.O. Box 15551, Al-Ain, United Arab Emirates \\
$^3$ Physical-Technical Institute,  Uzbek Academy of Sciences,
100084, Tashkent, Uzbekistan}

\date{\today}

\begin{abstract}
The existence regimes and dynamics of soliton molecules in
dispersion-managed (DM) optical fibers have been studied.
Initially we develop a variational approximation (VA) for
description of periodic dynamics of a soliton molecule within each
unit cell of the dispersion map. The obtained system of coupled
equations for the pulse width and chirp allows to find the
parameters of DM soliton molecules for the given dispersion map
and pulse energy. Then by means of a scaling transformation and
averaging procedure we reduce the original nonlinear Schr\"odinger
equation (NLSE) with piecewise-constant periodic dispersion to its
counterpart with constant coefficients and additional parabolic
potential. The obtained averaged NLSE with expulsive potential can
explain the essential features of solitons and soliton molecules
in DM fibers related to their energy loss during propagation.
Also, the model of averaged NLSE predicts the instability of the
temporal position of the soliton, which may lead to difficulty in
holding the pulse in the middle of its time slot. All numerical
simulations are performed using the parameters of the existing DM
fiber setup, and illustrated via pertinent examples.
\end{abstract}
\pacs{42.81.Dp, 42.65.Re, 42.79.Sz}

\maketitle

\section{Introduction}

During the last two decades the amount of information transmitted
via optical fiber communication systems has increased enormously.
High performance of modern communication lines is provided by an
optical fiber \cite{mitschke-book}, where the information is
encoded and transmitted as a sequence of light pulses. Despite the
very high data rate achieved today ($\sim$ 50 TB/s over single
fiber), there is urgent demand for even more capacity of the fiber
line, originating from the needs of telecommunications and the
Internet. Different approaches are being pursued to address the
capacity problem, such as improving the transmission properties of
the optical fiber, lying additional parallel cables and using the
information coding schemes going beyond the binary format, which
is currently employed. Among the above mentioned approaches using
the extended alphabet, where a bound state of two or more optical
solitons called {\it soliton molecule} serves as a new carrier of
information unit \cite{stratmann2005}, seems to be most appealing
because it employs already existing optical fiber lines, and
therefore is beyond the competition from the viewpoint of
cost-effectiveness.

It is appropriate to recall that solitons are self-localized wave
packets that can propagate along wave-guides preserving their
shape and velocity, and exhibit particle-like collisions with
each-other. Bright solitons emerge from the fine balance between
the dispersive broadening and nonlinear self-focusing of the wave
packet. When the optical soliton was theoretically predicted
\cite{hasegawa1973} and experimentally observed
\cite{mollenauer1980}, there was a promising idea that it can be
used as information carrier in optical fiber communication systems
\cite{hasegawa1995} due to its exceptional robustness against
perturbations. Besides, the nonlinearity of the optical fiber,
considered to be nuisance in linear systems, has been used as an
advantage in this case as it provides the soliton's self-healing
property. However, some other detrimental effects like four-wave
mixing and Gordon-Haus timing jitter \cite{mollenauer2006} have
imposed difficulties for the progress in this direction. Later the
concept of dispersion-management (DM) and DM soliton was put
forward (for a recent review see \cite{turitsyn2012}), which
allowed to suppress these adverse effects, and eventually has led
to realization of several commercial soliton based optical fiber
communication systems. An interesting chronicle of the growth of
the telecommunications industry, including the fiber optic
systems, is given in Ref. \cite{mitschke-book}.

Recent progress in using solitons for information transfer is
linked to the observation, that solitons in DM fibers can form
stable bound states called soliton molecules, and realizing their
potential for enhancing the capacity of the communication system
via extension of the coding alphabet \cite{stratmann2005}. The
binding mechanism of solitons in the molecule was proposed in
\cite{hause2008}. The proof-of-principle demonstration of the data
transmission via optical fiber using the extended alphabet:
logical {\it zero}, {\it one} (single soliton), {\it two}
(two-soliton molecule) and {\it three} (three-soliton molecule),
was recently reported in \cite{rohrmann2012,rohrmann2013}. The
practical implementation of this novel approach requires extensive
research on the existence regimes, stability, mutual interactions
and propagation dynamics of soliton molecules in DM fibers.

In this work we study the existence regimes and dynamics of
soliton molecules in DM fibers by analytical and numerical means.
First we develop a variational approach (VA) to find the
stationary shape of the molecule and equilibrium separation
between solitons in the molecule. At this stage we obtain two
coupled ordinary differential equations (ODE) for the temporal
separation between solitons and chirp parameter, which describes
the fast dynamics of the molecule within a unit cell of the DM
fiber. Although the derived ODE system is capable of describing
the propagation of the molecule for arbitrary distance, long-haul
transmission of DM solitons and molecules is convenient to explore
using the averaged nonlinear Schr\"odinger equation (NLSE). The
averaged NLSE can help to specify the existence regimes of
solitons and molecules, elucidate the physical mechanism by which
they loose energy, and eventually disintegrate in conservative DM
fibers.

The paper is organized as follows. In the next Sec. II we
introduce the NLSE which governs the pulse propagation in DM
fibers. Here we also present the parameters of the DM fiber used
in our calculations. In Sec. III we develop the VA for the fast
dynamics of solitons and molecules and compare its predictions
with the results of PDE simulations. Then we derive in Sec. IV the
averaged NLSE and determine its coefficients using the VA. Here we
also perform the analysis of the pulse propagation in the NLSE
with inverted parabolic potential. In Sec. V we summarize our
findings.

\section{The governing equation}

Propagation of optical pulses in fibers with inhomogeneous
parameters is described by the following nonlinear Schr\"odinger
equation
\begin{equation}\label{nlse0}
i\frac{\partial E}{\partial z}-\frac{\beta(z)}{2}\frac{\partial^2
E}{\partial t^2} + \Gamma(z)|E|^2 E = i g(z) E,
\end{equation}
where $E(z,t) \ (|E|^2 \ [W])$, $z \ [m]$, and $t \ [s]$ are
respectively, the complex envelope of the electric field, the
propagation distance, and the retarded time. The coefficients
$\beta(z) \ [s^2/m]$, $\Gamma(z) \ [1/(W \cdot m)]$, and $g(z) \
[1/m]$ represent the fiber's group velocity dispersion (GVD),
nonlinearity and gain/loss parameter, respectively. Here and below
in rectangular brackets $[ \cdot \cdot \cdot ]$ we show the
physical unit of the corresponding variable.

For qualitative analysis and numerical simulations it is
convenient to reduce the Eq. (\ref{nlse0}) into dimensionless
form. At first we eliminate the gain/loss term via new variable
$u(z,t) \ (|u|^2 \ [W])$, following Ref. \cite{maruta2002}
\begin{equation}\label{gain}
E(z,t) = a(z) u(z,t), \qquad a(z)=a_0 \exp\left[\int_0^z g(\xi)
d\xi\right],
\end{equation}
where $a_0$ is dimensionless constant. The new complex function
$u(z,t)$ satisfies the equation
\begin{equation}\label{nlse1}
i\frac{\partial u}{\partial z}-\frac{\beta(z)}{2}\frac{\partial^2
u}{\partial t^2} + \gamma(z)|u|^2 u = 0,
\end{equation}
where $\gamma(z) = a^2(z) \cdot \Gamma(z)$ is  the fiber's
effective nonlinearity. Now we convert the Eq. (\ref{nlse1}) into
standard form with constant nonlinearity by introducing the new
coordinate $z^{\prime} \ [1/W]$ defined by $ z^{\prime}(z) =
\int^{z}_0 \gamma(\xi) d\xi$
\begin{equation}\label{nlse2}
i\frac{\partial u}{\partial z^{\prime}}
-\frac{\beta^{\prime}(z^{\prime})}{2}\frac{\partial^2 u}{\partial
t^2} + |u|^2 u = 0,
\end{equation}
where $\beta^{\prime}(z^{\prime}) =
\beta(z^{\prime})/\gamma(z^{\prime})$ \ [W $\cdot$ s$^2$] is the
fiber's effective dispersion, characterizing both the fiber's GVD
and nonlinearity. The original parameters $\beta(z)$, $\Gamma(z)$
and $g(z)$ in Eq. (\ref{nlse0}) are periodic functions of
propagation distance with common period $L$, while the effective
dispersion $\beta^{\prime}(z^{\prime})$ in Eq. (\ref{nlse2}) is a
periodic function with period defined by
\begin{equation}\label{lprime}
L^{\prime} = \int^{L}_0 \gamma(\xi) d\xi = L^{+}\gamma^{+} +
L^{-}\gamma^{-}, \qquad [1/W].
\end{equation}
To obtain the final equation we introduce dimensionless variables
\begin{equation}\label{dim}
q(Z,T) = u(z^{\prime},t) \cdot \sqrt{L^{\prime}}, \quad Z =
z^{\prime}/L^{\prime} \quad T = t/\tau_m,
\end{equation}
where $\tau_m$ is the characteristic time scale equal to pulse
duration of the laser source $\tau_{fwhm}$. In terms of these
variables the dimensionless governing equation acquires the form
\begin{equation}\label{nlse3}
i\frac{\partial q}{\partial Z}-\frac{D(Z)}{2}\frac{\partial^2
q}{\partial T^2} + |q|^2q = 0,
\end{equation}
where $D(Z)= \beta^{\prime}(Z) L^{\prime}/\tau_m^2$ represents the
fiber's dimensionless effective dispersion. To show the results in
dimensional variables we solve the equation
\begin{equation}\label{z}
\int^{z}_0 \gamma(\xi) d\xi = Z \cdot L^{\prime},
\end{equation}
with respect to $z$ [m], for given dimensionless propagation
distance $Z$, the period $L^{\prime}$ [1/W], and known nonlinear
map function $\gamma(z)$ [$1/(W \cdot m)$]. The original time $t$
[s] and field amplitude $u$ $[\sqrt{W}]$ are restored via
Eq.(\ref{dim}). Note that $z=L$ corresponds to $Z=1$, in
accordance with Eq. (\ref{lprime}).

In the absence of gain/loss term in Eq. (\ref{nlse0}), i.e. $g(z)
=0$
\begin{equation}
a(z) \equiv a_0, \qquad a_0 = \left( \frac{L}{\int^L_0 \exp[2
\cdot \int^z_0 g(\xi) d \xi] \, dz} \right)^{1/2} =1.
\end{equation}
The dimensionless pulse energy is
\begin{equation}\label{energy}
E_0 = \int \limits_{-\infty}^{\infty} |q|^2 dT =
\frac{L^{\prime}}{\tau_m} \cdot \int \limits_{-\infty}^{\infty}
|u|^2 dt = \frac{L^{\prime}}{\tau_m} \cdot {\cal E}_0,
\end{equation}
where ${\cal E}_0$ \ [J] - is the original pulse energy.

\subsection{Parameters of the dispersion map}\label{params}

\begin{figure}[htb]
\centerline{
\includegraphics[width=8cm,height=4cm]{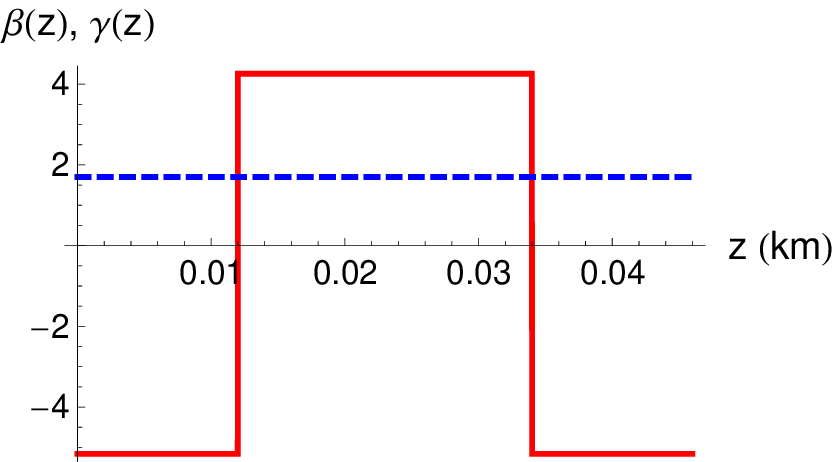} \qquad
\includegraphics[width=8cm,height=4cm]{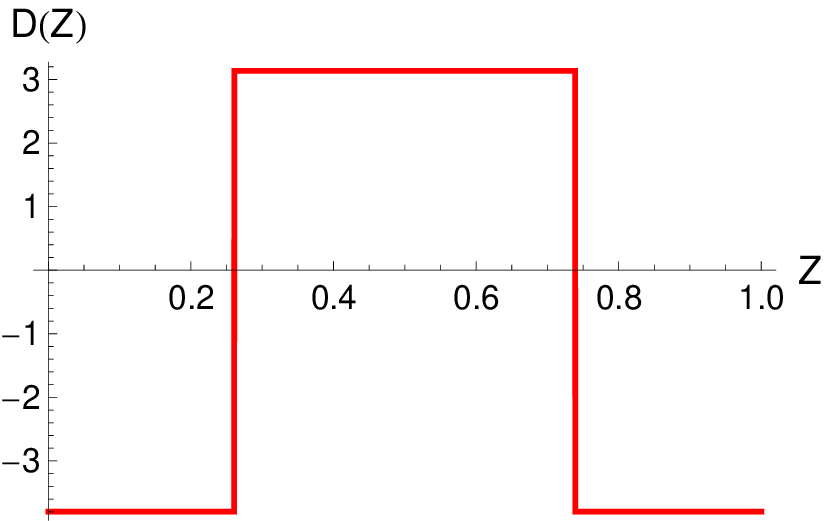}}
\caption{Dispersion profile for one period of the DM map. Red
solid line represents the piecewise-constant dispersion, blue
dashed line is the parameter of nonlinearity. Left panel: For the
original Eq. (\ref{nlse1}). Right panel: For reduced dimensionless
Eq.~(\ref{nlse3}). Note that here the coefficient of nonlinearity
is equal to one and has not been shown in Eq. (\ref{nlse3}) and
figure.} \label{fig1}
\end{figure}

In the following sections we employ the DM map parameters,
corresponding to the setup of Ref.
\cite{rohrmann2012,rohrmann2013}, for laser wavelength
$\lambda=1540$ nm
\begin{itemize}
\item[] $\beta^{+}_2$ = $+$ 4.259 [ ps$^2$/km ], \quad
$\gamma^{+}$=1.7 [ 1/(W $\cdot$ km) ], \quad $L^{+}$ = 22 [ m ]
(positive GVD fiber),
\item[] $\beta^{-}_2$ = $-$ 5.159 [ ps$^2$/km ], \quad $\gamma^{-}$=1.7 [ 1/(W $\cdot$ km) ], \quad
$L^{-}$ = 24 [ m ] (negative GVD fiber).
\end{itemize}
The period of the original DM map is $L = L^{+} + L^{-}$ = 46 [ m ].
For the effective dispersion in Eq. (\ref{nlse2}) the period is
$L^{\prime} = \gamma^{+} \cdot L^{+} + \gamma^{-} \cdot L^{-}$ = 0.078 [ 1/W ]. The
path averaged dispersion and nonlinearity are equal to
\begin{itemize}
\item[] $\bar{\beta}_2 = (\beta^{+}_2 \cdot L^{+} + \beta^{-}_2 \cdot L^{-})/L$ =
$-$ 0.655 [ ps$^2$/km ], \quad $\bar{\gamma} = (\gamma^{+} \cdot L^{+} + \gamma^{-}
\cdot L^{-})/L$ = 1.7 [ 1/(W $\cdot$ km) ].
\end{itemize}
The characteristic time, length and energy scales are given below
\begin{itemize}
\item[] $\tau_{fwhm}=0.25$ [ ps ], \quad $T_0 =\tau_{fwhm}/1.67 = 0.15$ [ ps ] (pulse duration),
\item[] $L_D =T_0^2/|\bar{\beta}_2| = 0.034$ [ km ] (dispersion length),
\item[] $S = \left(|\beta^{+}_2-\bar{\beta}_2|\cdot L^{+} +
|\beta^{-}_2-\bar{\beta}_2| \cdot L^{-} \right)/\tau^2_{fwhm}$ = 3.459 (strength of the map).
\end{itemize}
From the above presented data we get the dimensionless map parameters
\begin{itemize}
\item[] $D_1 = (\beta^{-}_2/\gamma^{-}) \cdot
(L^{\prime}/\tau_{fwhm}^2)$ = $-$ 3.796, \quad
        $L_1 = L^{-} \cdot \gamma^{-} / L^{\prime}$ = 0.522,
\item[] $D_2 = (\beta^{+}_2/\gamma^{+}) \cdot
(L^{\prime}/\tau_{fwhm}^2)$ = $+$ 3.135,  \quad
        $L_2 = L^{+} \cdot \gamma^{+} / L^{\prime}$ = 0.478,
\item[] $\Delta D = (D_1 \cdot L_1 + D_2 \cdot L_2)/(L_1 + L_2)$ =
$-$ 0.482, \quad $L_1 + L_2$ = 1.
\end{itemize}
Figure \ref{fig1} shows the dispersion profiles for the original
and reduced governing equations.

\section{The variational approach for fast dynamics of soliton
molecules}\label{secva}

Propagation of a DM soliton is characterized by fast variation of
the pulse shape within each period of the DM map and slow
variations on longer distances when observed stroboscopically,
i.e. once per dispersion period. The periodic dynamics is compromised
by different imperfections of the fiber, continuous emission of
linear waves by the soliton and non-perfect initial shape of the
DM soliton injected into the fiber. To compensate for the power
loss and other distortions, in practice the pulses are regenerated
by amplifiers which are equidistantly installed along the fiber
line.

The variational approach, initially developed for description of
optical soliton propagation in homogeneous fibers
\cite{anderson1983}, later was successfully applied to DM solitons
\cite{turitsyn1998} and antisymmetric solitons in DM fibers
\cite{pare1999,feng2004}. The advantages and disadvantages of the VA
as compared to other methods of exploring DM solitons are
discussed in \cite{cautaerts2000,malomed-book}. Below we elaborate
the VA for two- and three-soliton molecules in DM fibers, based on
the reduced NLSE (\ref{nlse3}).

\begin{figure}[htb]
\centerline{
\includegraphics[width=8cm,height=6cm]{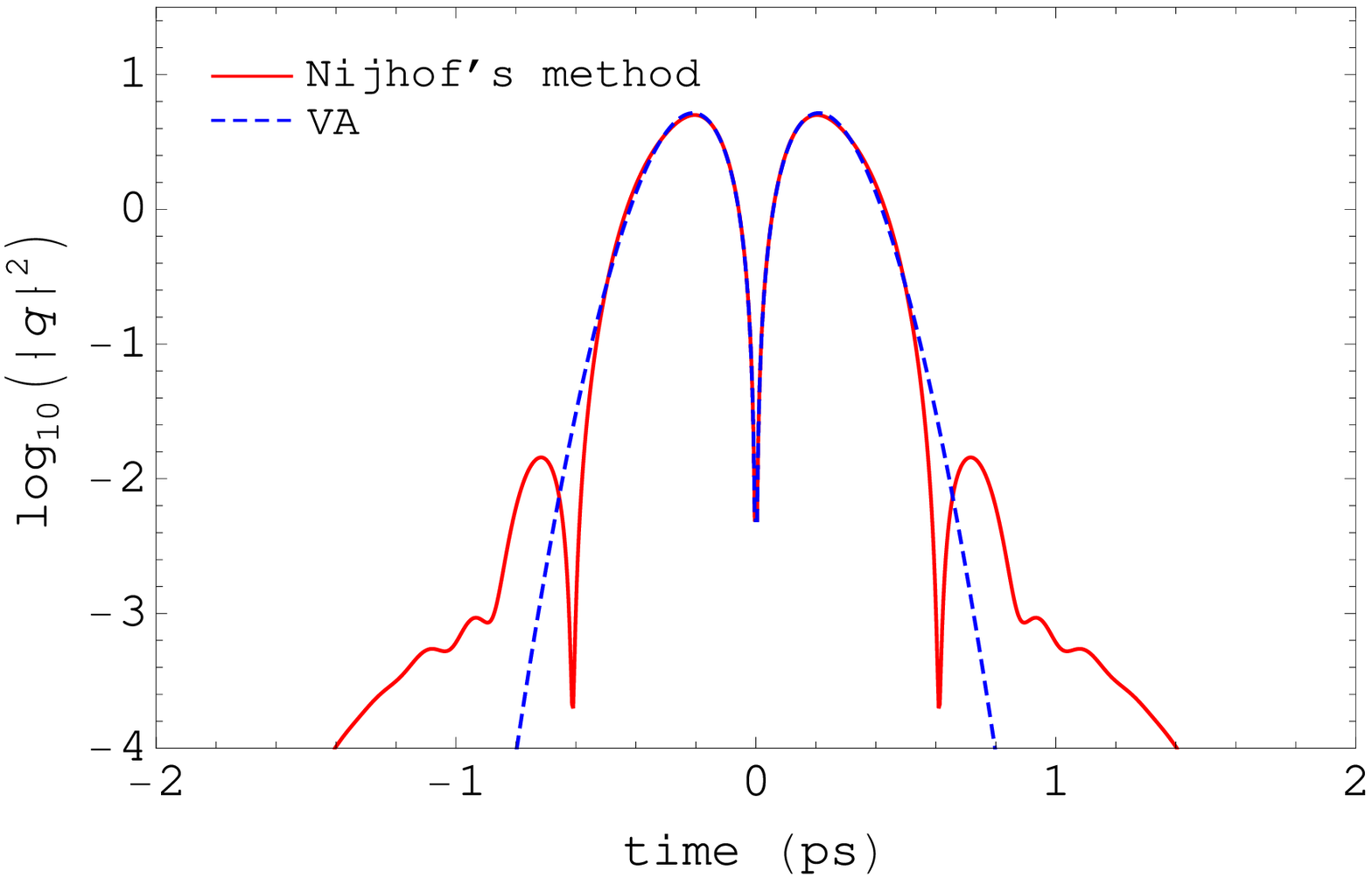} \qquad
\includegraphics[width=8cm,height=6cm]{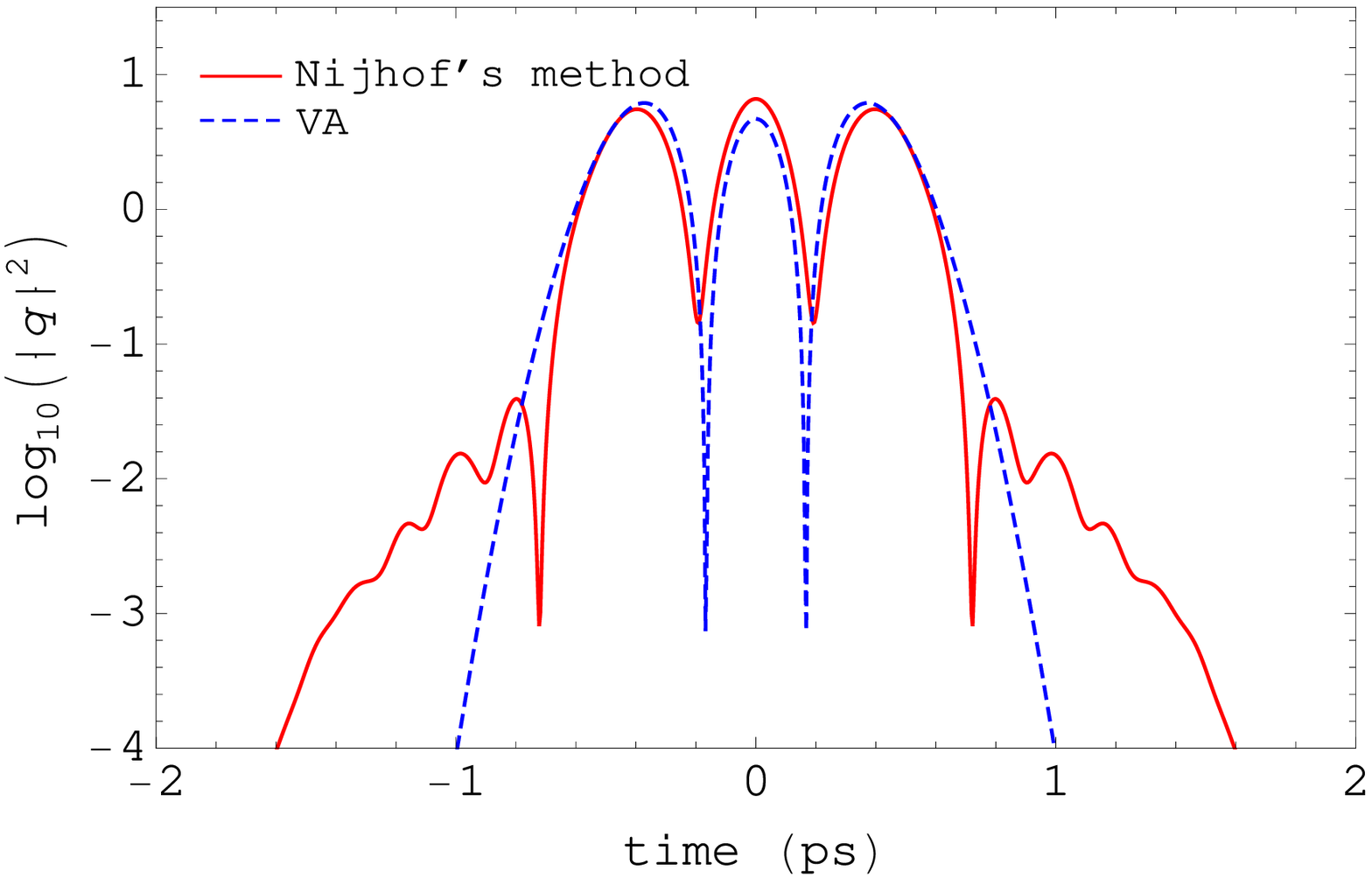}}
\caption{Pulse shapes of the two-soliton molecule (left panel) and
three-soliton molecule (right panel) in logarithmic scale. Red
solid lines correspond to waveforms found by Nijhof's method
\cite{nijhof2000} applied to original NLSE (\ref{nlse3}), while
blue dashed lines are the prediction of VA equations
(\ref{tau})-(\ref{alpha}). Appreciable deviations of the two
waveforms are seen only in far tails, where the pulse amplitude is
less than a percent of the maximum value. The parameters for the
2-soliton molecule: $E_0 = 10.448$ (in original units ${\cal E}_0
= 20$ pJ), $\tau_0 = 1.1828$, $A_0= \frac{2}{\tau_0} \,
\sqrt{\frac{E_0}{\tau_0} \, \sqrt{2/\pi}}=4.489$. For the
3-soliton molecule: $E_0 = 15.672$ (${\cal E}_0 = 30$ pJ), $\tau_0
= 1.333$, $A_0=\sqrt{\frac{E_0}{\tau_0 \sqrt{2\pi}}}=2.166$. }
\label{fig2}
\end{figure}

\begin{figure}[htb]
\centerline{
\includegraphics[width=8cm,height=6cm,clip]{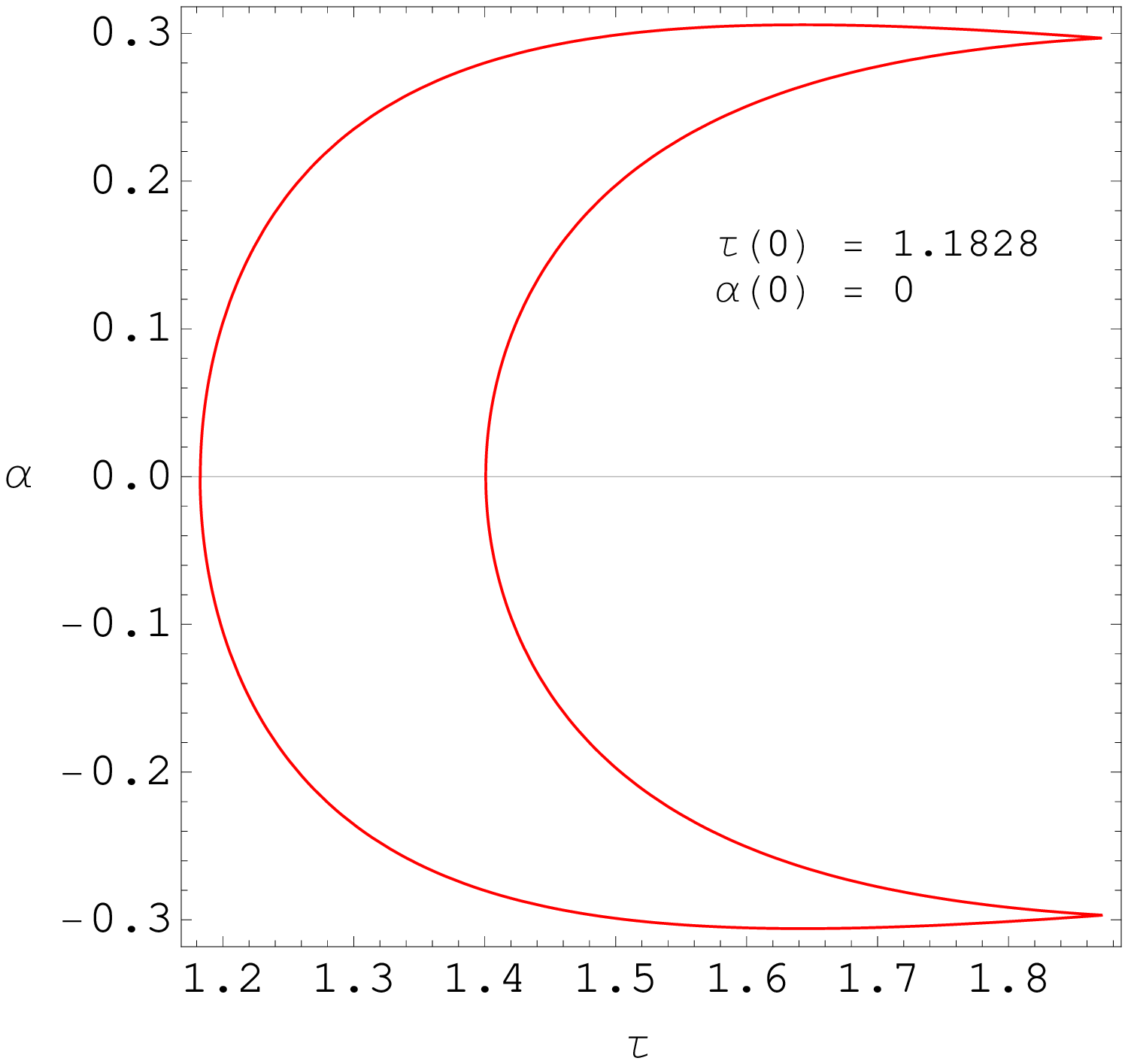} \qquad
\includegraphics[width=8cm,height=6cm,clip]{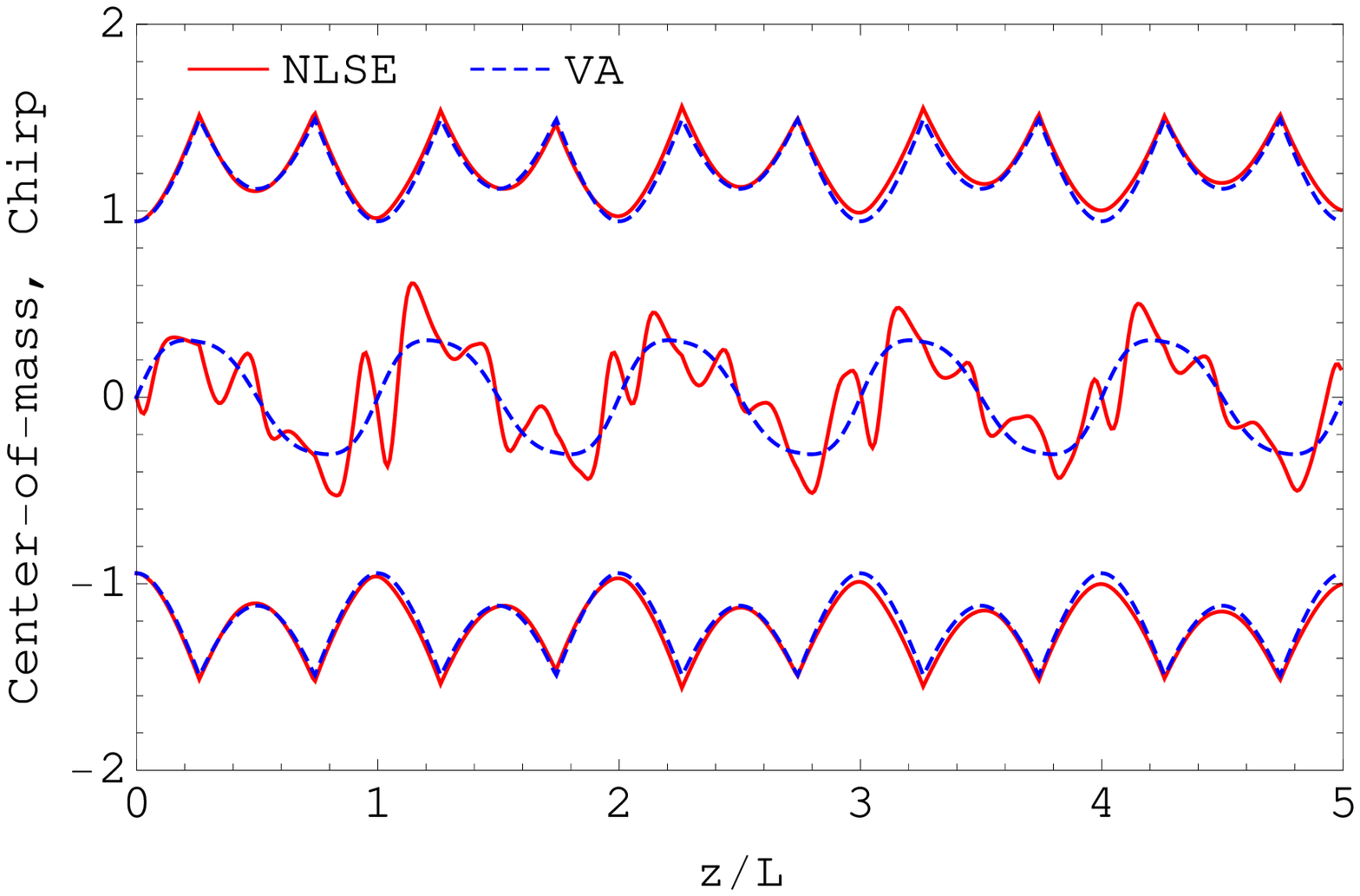}}
\caption{Left panel: Closed phase trajectory indicates that the
fixed point $\tau(0)=1.1828$, $\alpha(0)=0$ corresponds to the
periodic solution of the system (\ref{tau})-(\ref{alpha}) for a
two-soliton molecule with energy $E=10.448$, shown in
Fig.\ref{fig2}. Right panel: The solution of the NLSE
(\ref{nlse3}) with the initial pulse shape predicted by the VA
shows that solitons in the two-soliton molecule perform
oscillatory motion periodically reducing and increasing the
temporal separation between their center-of-mass positions (top
and bottom curves). The evolution of the chirp parameter (middle
curve) shows qualitative agreement with the VA.} \label{fig3}
\end{figure}

The Eq. (\ref{nlse3}) can be obtained from the following
Lagrangian density (for tidier notations we again use small
letters instead of capitals)
\begin{equation}\label{lagr}
{\cal L} = \frac{i}{2}\left(q
q_z^{\ast}-q^{\ast}q_z\right)-\frac{d(z)}{2}|q_t|^2 -
\frac{1}{2}|q|^4.
\end{equation}
To derive the VA equations we consider the shape of the soliton
molecule in general form \cite{turitsyn1998,turitsyn1997}
\begin{equation}\label{ansatz}
q(z,t)=\frac{1}{\tau^{1/2}} \cdot f(t/\tau) \cdot e^{i\alpha t^2 +
i \sigma},
\end{equation}
where $\alpha(z), \sigma(z)$ are the chirp parameter and phase,
$\tau(z)$ is the pulse duration (proportional to the separation
between solitons in the molecule), $f(x)$ is the real function
which represents the stationary shape of the soliton molecule. The
energy of the pulse is (subscript zero in Eq. (\ref{energy}) is
dropped)
\begin{equation}
E = \int_{-\infty}^{\infty} |q|^2 dt = \int_{-\infty}^{\infty}
f^2(x) dx, \qquad x=t/\tau.
\end{equation}
Substitution of the trial function (\ref{ansatz}) into Eq.
(\ref{lagr}) yields the Lagrangian density
\begin{equation}
{\cal L} = \tau \alpha_z x^2 f^2 + \frac{\sigma_z}{\tau} f^2 -
\frac{1}{2}\frac{d(z)}{\tau^3} f_x^2 -
2\,d(z)\,\tau\,\alpha^2\,x^2\,f^2 -\frac{1}{2\tau^2} \, f^4.
\end{equation}
The averaged Lagrangian is obtained by integrating the last
expression over the reduced time variable
\begin{equation}\label{lagrdensity}
L = \alpha_z \tau^2 \int_{-\infty}^{\infty} x^2\,f^2 dx + \sigma_z
\int_{-\infty}^{\infty} f^2 dx - \frac{d(z)}{2\tau^2}
\int_{-\infty}^{\infty}f_x^2 dx - 2d(z)\alpha^2 \tau^2
\int_{-\infty}^{\infty} x^2 f^2 dx - \frac{1}{2\tau}
\int_{-\infty}^{\infty}f^4 dx,
\end{equation}
with a few integral constants, determined solely by the pulse
shape $f(x)$. The Euler-Lagrange equations with respect to
variational parameters $\tau, \alpha, \sigma$ give rise to a
coupled set of ODE's
\begin{eqnarray}
\tau_z   &=& -2 \, d(z) \,\alpha \, \tau, \label{tau} \\
\alpha_z &=& -2 \, d(z) \left( \frac{c_1}{4 \tau^4} - \alpha^2
\right) - \frac{c_2}{4\, \tau^3}, \label{alpha}
\end{eqnarray}
where
\begin{equation}\label{c1c2}
c_1=\int^{\infty}_{-\infty}f_x^2 dx \, \mbox{\Large /}
\int^{\infty}_{-\infty} x^2\,f^2 dx, \qquad
c_2=\int^{\infty}_{-\infty}f^4 dx   \, \mbox{\Large /}
\int^{\infty}_{-\infty} x^2\,f^2 dx.
\end{equation}
We adopt the following trial functions $f(x)$ to specify the
shapes of the pulses and define the corresponding parameters
\begin{eqnarray}
\mbox{single soliton}&:& f(x)=A \, e^{-x^2}, \quad E = A^2 \cdot
\sqrt{\frac{\pi}{2}}, \quad \ c_1 = 4, \quad
c_2 = \frac{4 \cdot E}{\sqrt{\pi}}. \label{1sol} \\
\mbox{2-soliton molecule}&:& f(x)=A \, x \, e^{-x^2}, \quad \ E =
\frac{A^2}{4}\cdot \sqrt{\frac{\pi}{2}}, \quad \ c_1 = 4, \quad
c_2 = \frac{E}{\sqrt{\pi}}. \label{2sol} \\
\mbox{3-soliton molecule}&:& f(x)=A \, (4 x^2 -1) \, e^{-x^2},
\quad E = A^2 \sqrt{2\pi}, \quad c_1 = 4, \quad c_2 =
\frac{41\,E}{80\,\sqrt{\pi}}. \label{3sol}
\end{eqnarray}
The fixed point ($\tau_0, \alpha_0$) of the coupled system
(\ref{tau})-(\ref{alpha}), which can be found using the Nijhof's
method for VA models \cite{nijhof2000}, defines the stationary
shapes of the pulses with given energy $E$ and DM map function
$d(z)$. In Fig. \ref{fig2} we illustrate the shapes of the
two-soliton and three-soliton molecules, which are found by
solving the VA system (\ref{tau})-(\ref{alpha}) and compare with
the corresponding results of the Nijhof's method applied to the
original Eq. (\ref{nlse3}).
In Fig. \ref{fig3} we show the phase trajectory for the periodic
solution of the VA system (\ref{tau})-(\ref{alpha}) and temporal
center-of-mass positions of solitons in the two-soliton molecule
while it propagates along the fiber. The center-of-mass position
of the right/left soliton is calculated from the solution of the
VA system (\ref{tau})-(\ref{alpha}) as $\tau_{cm}(z) = \pm
\tau(z)\sqrt{2/\pi}$, while for the NLSE (\ref{nlse3}) the
corresponding formula is
\begin{equation}\label{cm}
\tau_{cm}(z)=\pm \ \frac{2}{E_0}\int^{\infty}_0 t \ |q(z,t)|^2 dt,
\end{equation}
where in actual calculations we use the temporal domain's half
length as the upper limit of the integration.
For evaluation of the pulse width and chirp parameter from the
solution of Eq. (\ref{nlse3}) we employ the following
expressions~\cite{nijhof2000}.
\begin{equation}
\tau(z)=\left(\frac{4}{3 E}\int^{\infty}_{-\infty}t^2 \cdot
|q(z,t)|^2 dt \right)^{1/2}, \quad
\alpha(z)=\frac{\int^{\infty}_{-\infty}{\rm Im}\left(q^2(z,t)
\cdot q^{\ast}_t(z,t) \right) dt}{\int^{\infty}_{-\infty}
|q(z,t)|^4 dt}.
\end{equation}
As can be seen from Fig. \ref{fig3} the dynamics of the pulse
width is described by the VA equations quite accurately, while for
the chirp parameter the agreement is only qualitative. It should
be noted that although the chirp parameter shows complicated
behavior within the map period, ``zero chirp" condition in the
middle of each anomalous GVD fiber is well satisfied.

\section{Averaged equation for soliton molecules in DM fibers}

Long-haul propagation of soliton molecules in DM fibers is
convenient to study using the averaged NLSE. The averaging
procedure was developed in \cite{turitsyn1997,turitsyn1998}. Below
we use the approach based on the scaling arguments proposed in
Ref. \cite{kodama1997} and look for the solution of Eq.
(\ref{nlse3}) in the form (capital letters changed to lower-case)
\begin{equation}
q(z,t)=w(z,t)\,e^{i\, \alpha(z)\,t^2},
\end{equation}
where $\alpha(z)$ is the chirp parameter. Inserting this into the
governing Eq. (\ref{nlse3}) we obtain
\begin{equation}\label{eq2}
i\left(w_z - 2\, d \, \alpha \, t \, w_t  \right) - \frac{d}{2}
w_{tt} + (2 d \alpha^2 - \alpha_z) \, t^2 w + |w|^2 w = i d \alpha w.
\end{equation}
By introducing the new time ($x$) and amplitude ($b$) functions
\begin{equation}\label{eq3}
x = t/\tau(z), \qquad w(z,t) = b(z)\, v(z,x).
\end{equation}
the last equation can be reduced into the form
\begin{equation}\label{eq4}
i \left(b v_z + b_z v - \frac{b t \tau_z}{\tau^2} v_x - \frac{2 \,
b \, d \, \alpha \, t}{\tau} v_x \right) -
\frac{b\,d}{2\,\tau^2}\, v_{x x} - (\alpha_z -2 \, d\,\alpha^2) \,
t^2 \, b \, v + b^3 |v|^2 v = i\, d \, \alpha \, b \, v.
\end{equation}
When the following relations are satisfied
\begin{equation}\label{eq5}
b_z = d \, \alpha \, b,  \qquad \tau_z = - 2 \, d \, \alpha \,
\tau,
\end{equation}
the NLSE with a parabolic potential results from Eq. (\ref{eq4})
\begin{equation}\label{eq6}
i v_z - \frac{d}{2 \, \tau^2} \, v_{x x} + b^2 |v|^2 v - (\alpha_z
- 2 d\, \alpha^2) \, \tau^2\, x^2 \, v = 0.
\end{equation}
The amplitude function is linked to the pulse width as $b(z) =
\beta/\sqrt{\tau (z)}$, which can be readily verified from Eq.
(\ref{eq5}). The constant $\beta$ is specified by the selected
trial function. Now applying the averaging procedure to Eq.
(\ref{eq6}) we obtain in leading order
\begin{equation}\label{averaged}
i\psi_z + d_0 \, \psi_{xx} + b_0 \, |\psi|^2 \psi + k_0 \, x^2 \,
\psi = 0,
\end{equation}
where $\psi(z,x)$ is the slowly varying core of the DM soliton.
The quantities averaged over one dispersion period are defined as
$d_0= -\frac{1}{2}\left<d(z)/\tau^2(z)\right>$, $b_0 = \beta^2
\left<1/\tau(z)\right>$, $k_0 = \left< [2 d(z) \, \alpha^2(z) -
\alpha_z(z)] \, \tau^2(z) \right>$. The averaging is performed as
$<\phi(z)>=\int \limits_0^1 \phi(z) \tau(z) dz /\int \limits_0^1
\tau(z) dz$ (for the reduced map period $L=1$). In numerical
implementation of the averaging procedure we employ the periodic
solution of VA Eqs. (\ref{tau})-(\ref{alpha}). In particular from
Eq. (\ref{alpha}) it follows that $k_0 = (c_1/2) \,
\left<d(z)/\tau^2(z)\right> + (c_2/4)\,\left<1/\tau(z)\right>$,
with constants $c_1, c_2$ given by Eq. (\ref{c1c2}). For the DM
map parameters specified in Sec. \ref{params} all coefficients of
Eq. (\ref{averaged}) appear to be positive ($d_0
> 0$, $b_0 > 0$, $k_0 > 0$), therefore we have gotten the NLSE with
anomalous dispersion, focusing nonlinearity and inverted parabolic
potential. This equation is formally similar to the quantum
mechanical equation for a wave packet, evolving under the
effective potential
\begin{equation}\label{potential}
U(z,x) = - b_0 \, |\psi(z,x)|^2 - k_0 \, x^2.
\end{equation}
The stationary solution $\psi(z,x)=\varphi(x) e^{i\lambda z}$ can
be found from the initial value problem with respect to variable
$x$
\begin{equation}\label{ivp}
d_0 \, \varphi_{xx} + b_0 \, \varphi^3 + k_0 \, x^2 \, \varphi -
\lambda \, \varphi = 0,
\end{equation}
and suitable initial conditions $\varphi(0)$ and $\varphi_x(0)$.

It as appropriate to mention that Eq. (\ref{averaged}) with
additional gain/loss term was previously considered also in other
contexts, such as the nonlinear compression of chirped optical
solitary waves \cite{moores1996}, and with regard to integrability
issues \cite{nakkeeran2001,mak2005,xu2003,grimshaw2007}.

\subsection{The variational approach for averaged NLSE}

To study the evolution of pulses governed by Eq. (\ref{averaged})
we develop the VA. The corresponding Lagrangian density is
\begin{equation}\label{lagr}
{\cal L} =
\frac{i}{2}(\psi\psi^*_z-\psi^*\psi_z)+d_0 |\psi_x|^2 -k_0 x^2 |\psi|^2 -\frac{b_0}{2}|\psi|^4.
\end{equation}
The trial function will be of the form
\begin{equation}\label{ansatz1}
\psi(x,t)= A \,\eta(x)\, e^{-(x-\xi)^2/\tau^2 + i\alpha (x-\xi)^2
+ iv(x-\xi)+i\varphi},
\end{equation}
where $A(z)$, $\tau(z)$, $\xi(z)$, $v(z)$, $\alpha(z)$,
$\varphi(z)$ are variational parameters, designating the pulse
amplitude, width, center of mass position, velocity, chirp and
phase, respectively. The auxiliary function $\eta(x)$ is
introduced for convenience and defines the type of the pulse.
Specifically for a single soliton $\eta(x)=1$, for two-soliton
molecule $\eta(x)=x$ and for three-soliton molecule $\eta(x)=4 x^2 - 1$.

\begin{figure}[htb]
\centerline{
\includegraphics[width=8cm,height=6cm,clip]{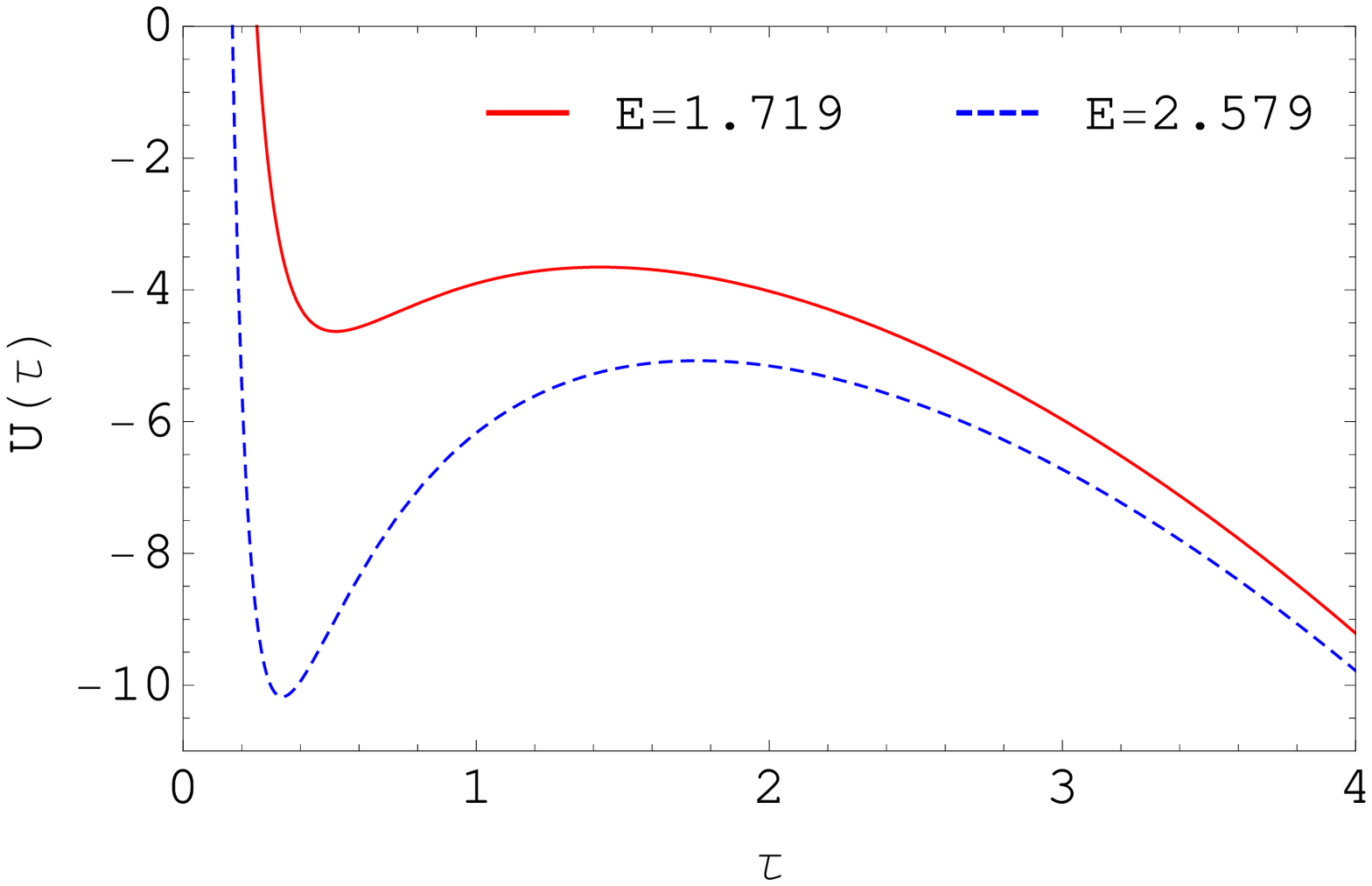} \qquad
\includegraphics[width=8cm,height=6cm,clip]{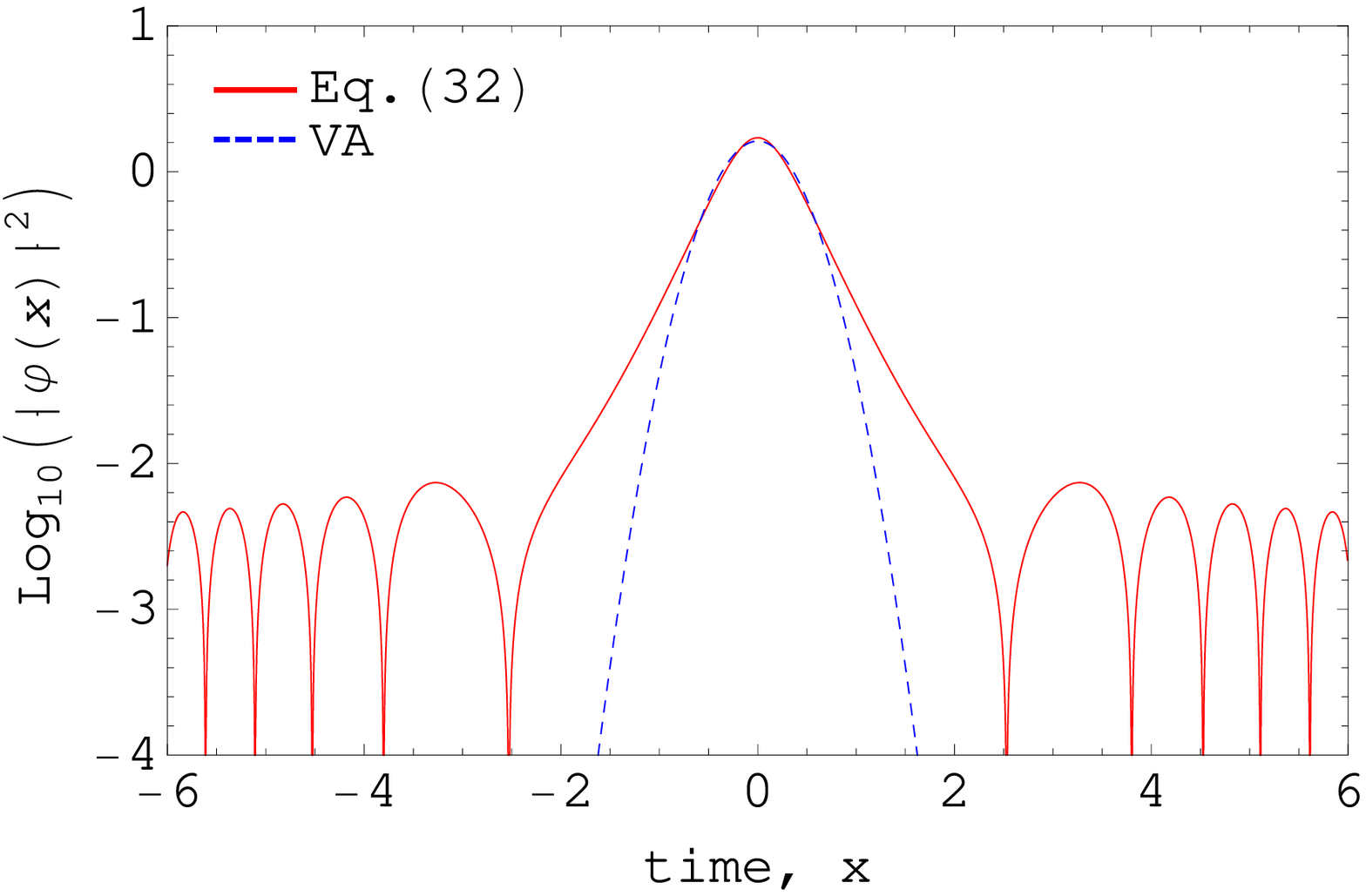}}
\caption{Left panel: The potential (\ref{pot1}) in the effective
particle model of VA for two values of the pulse energy. The pulse
with greater energy forms deeper potential well. The stationary
amplitude and width of the soliton for $E=1.7193$, according to VA
are $A=1.6227$ and $\tau=0.521$ respectively. The existence of a
local maximum and decay at large $\tau$ is the evidence of
meta-stability of the stationary state. Right panel: Comparison
between the pulse shapes (in logarithmic scale) obtained by
solving the stationary state problem (\ref{ivp}) for
$\lambda=4.6132$ and prediction of VA for the same pulse energy.
The wavy tails are due to the interference of waves reflecting
from the borders of the effective potential (\ref{potential}).
Parameters $d_0=0.3782$, $b_0=3.0917$, $k_0=0.6734$ are obtained
using the VA Eqs. (\ref{tau})-(\ref{alpha}) for a single DM
soliton with energy $E_0=5.224$ (in original units ${\cal E}_0=10$
pJ) and fiber data of Sec. \ref{params}. } \label{fig4}
\end{figure}

It is instructive to start with considering the existence and
dynamics of a single soliton on top of an inverted parabolic
potential. The integration $L=\int_{-\infty}^{\infty}{\cal L} dx$
using the trial function (\ref{ansatz1}) with $\eta(x)=1$ gives
rise to the following effective Lagrangian
\begin{equation}  \label{efflagr}
\frac{L}{E} = \frac{1}{4}\tau^2 \alpha_z -\xi_z^2 + \varphi_z +
\frac{d_0}{\tau^2} + d_0  \tau^2 \alpha^2 + d_0 \xi_z^2 -
\frac{k_0}{4}\tau^2 - k_0 \xi^2 - \frac{b_0 E}{2 \sqrt{\pi}\tau}.
\end{equation}
where the pulse energy $E=A^2 \tau \sqrt{\pi/2}$ is conserved. Now
applying the Euler - Lagrange equations with respect to the
variational parameters, we get the ODE system for the pulse width
and its center-of-mass position
\begin{eqnarray}
\tau_{zz} &=& \frac{16 \, d_0^2}{\tau^3} + 4 \, d_0 \, k_0 \, \tau
- \frac{4 \, d_0 \, b_0 \, E}{\sqrt{\pi}\, \tau^2}, \label{tau1} \\
\xi_{zz} &=& -k_0/(d_0-1) \, \xi.
\end{eqnarray}
The system for the two-soliton and three-soliton molecules will be
similar, with only re-scaled energy coefficient, $E\rightarrow
E/4$ for the former case and $E \rightarrow (41/320) E$ for the
latter case (see the coefficient $c_2$ in
Eqs.(\ref{2sol})-(\ref{3sol})). As can be seen from this system,
the center-of-mass and internal dynamics of the soliton are
decoupled. This is due to the property of a parabolic potential
and is the manifestation of the Ehrenfest's theorem (its validity
for the nonlinear Schr\"odinger equation with a linear and
parabolic potentials was proved in Ref. \cite{hasse1982}). In
other types of potentials these two degrees of freedom are
coupled. In fact the equation for the center-of-mass is the
harmonic oscillator equation with a purely imaginary frequency
$\omega^2= k_0/(d_0 - 1) < 0$, since in practical situations $k_0
> 0$ and $d_0 < 1$. Therefore, the center-of-mass of the soliton
is unstable against sliding down an inverted parabola with
exponentially increasing distance from the origin
\begin{equation}\label{com}
\xi(z)=\xi(0)\,e^{K z}, \quad \mbox{where} \quad K =
\sqrt{k_0/|d_0-1|} > 0.
\end{equation}
The equation for the pulse width (\ref{tau1}) is similar to the
equation of motion for a unit mass particle in the anharmonic potential
\begin{equation}\label{pot1}
\tau_{zz}=-\frac{d U}{d \tau}, \qquad
U(\tau)=\frac{8 \, d_0^2}{\tau^2} -2 \, d_0 \, k_0 \, \tau^2
- \frac{4 \, d_0 \, b_0 \, E}{\sqrt{\pi}\, \tau},
\end{equation}
which is depicted in Fig. \ref{fig4}. The minimum of this
potential is found from the solution of the quartic equation
\begin{equation}\label{quartic}
\tau^4-m \tau + n = 0, \quad m=\frac{b_0 E}{\sqrt{\pi}\, k_0},
\quad n=\frac{4 d_0}{k_0},
\end{equation}
and corresponds to the stationary width of the soliton. For a
given pulse energy $E$ this defines the shape of the soliton
according to the ansatz (\ref{ansatz1}). At some critical energy
$E_{cr}$ the local minimum in this potential disappears, which
means that Eq. (\ref{averaged}) does not support solitons and
molecules with energy below $E_{cr}$. The value of the critical
energy can be found from the condition that Eq. (\ref{quartic})
has a real solution, which takes place if $m > 4 (n/3)^{3/4}$, or
in terms of energy
\begin{equation}\label{ecrit}
E > E_{cr} = \frac{2^{7/2} \, \pi^{1/2}}{3^{3/4}} \cdot
\frac{d_0^{3/4} \, k_0^{1/4}}{b_0} \simeq 8.8 \, \frac{d_0^{3/4}
\, k_0^{1/4}}{b_0}.
\end{equation}
In fact the shape of the potential $U(\tau)$ gives the evidence
that solitons and molecules in the system are meta-stable. In the
same figure \ref{fig4} we compare the pulse profiles obtained by
solution of the stationary state Eq. (\ref{ivp}) with the
prediction of VA. The parameter $\lambda$ for Eq. (\ref{ivp}) is
obtained using the shooting method, where the pulse energy $E=\int
|\varphi(x)|^2 dx$ is minimized and initial conditions $\varphi(0)
= A$, $\varphi_x(0)=0$ are used. As can be seen from this figure,
the deviation between the two wave profiles is notable at far
tails of the pulse, where the field intensity significantly
decreases. The wavy tails on the numerically exact pulse profile,
found from solution of Eq. (\ref{ivp}) indicate the existence of
waves reflected from (and partially transmitted through) the
borders of the effective potential (\ref{potential}). The waves
escaping the effective potential contribute to the continuous
outflow of energy from the soliton propagating along the fiber,
and that will be the fundamental source of its instability in
conservative DM fibers.

\begin{figure}[htb]
\centerline{
\includegraphics[width=8cm,height=6cm,clip]{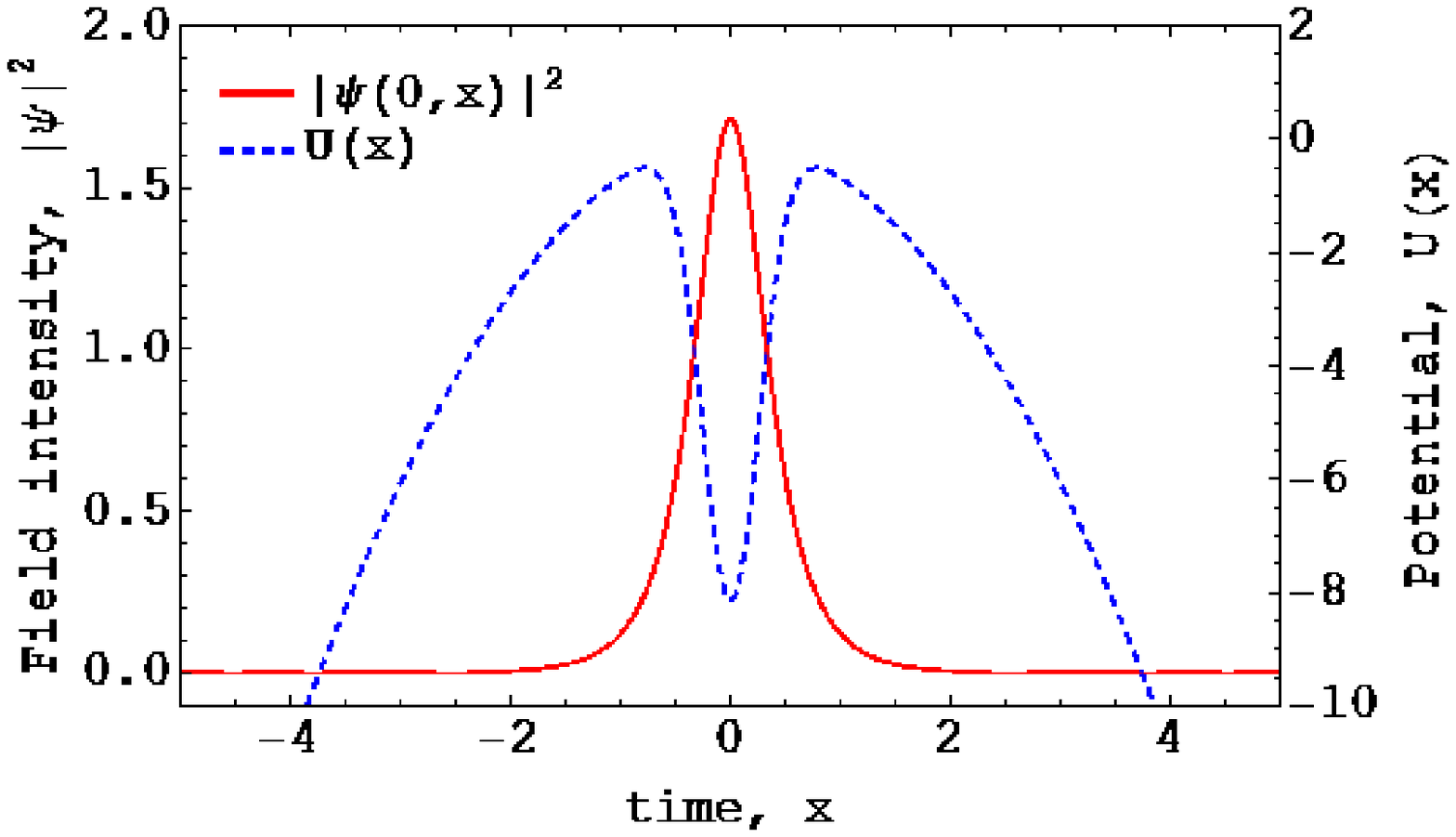} \qquad
\includegraphics[width=8cm,height=6cm,clip]{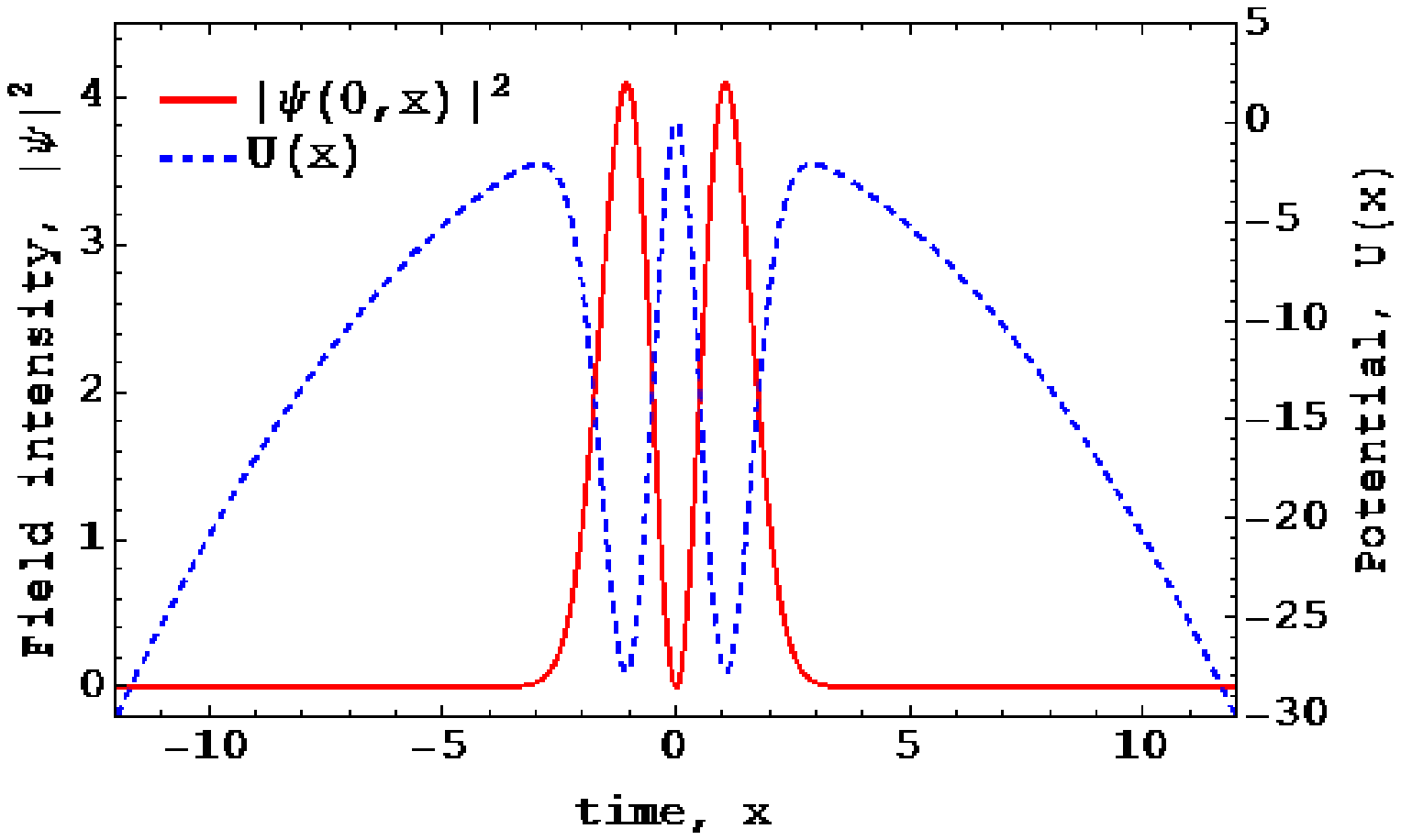}}
\caption{Single soliton (left panel) and two-soliton molecule
(right panel) of the averaged NLSE (\ref{averaged}) are shown by
red solid lines. The corresponding effective potentials according
to Eq. (\ref{potential}) are shown by blue dashed lines. The
parameters for the single soliton are the same as in previous
figure, while for the two-soliton molecule the parameters are
$d_0=0.1978$, $b_0=6.7274$, $k_0=0.209$, $E_0=10.448$ (in original
units ${\cal E}_0=20$ pJ). } \label{fig5}
\end{figure}

The ``pulse in the effective potential" picture, shown in Fig.
\ref{fig5}, is helpful for the analysis of instability issues. The
first aspect to be noted is that, the height and width of the
effective potential barrier in both directions from the pulse are
finite, and depends on the intensity of the pulse itself.
Therefore a continuous and nonlinearly progressing energy outflow
from the soliton takes place via the tunneling effect, whose rate
can be estimated by means of the semiclassical WKB method as done
for the matter-wave analogue of this problem in Ref.
\cite{carr2002}. These authors also have shown that the rate of
energy outflow (number of particles for condensates) nonlinearly
increases, eventually leading to disintegration of the soliton.
The second aspect is the instability of the center-of-mass
position of the soliton against sliding down an inverted parabola.
According to Eq. (\ref{com}) any small departure of the
center-of-mass position from the origin (the top of the inverted
parabola) in either direction will exponentially grow.

\begin{figure}[htb]
\centerline{\includegraphics[width=7cm,height=6cm,clip]{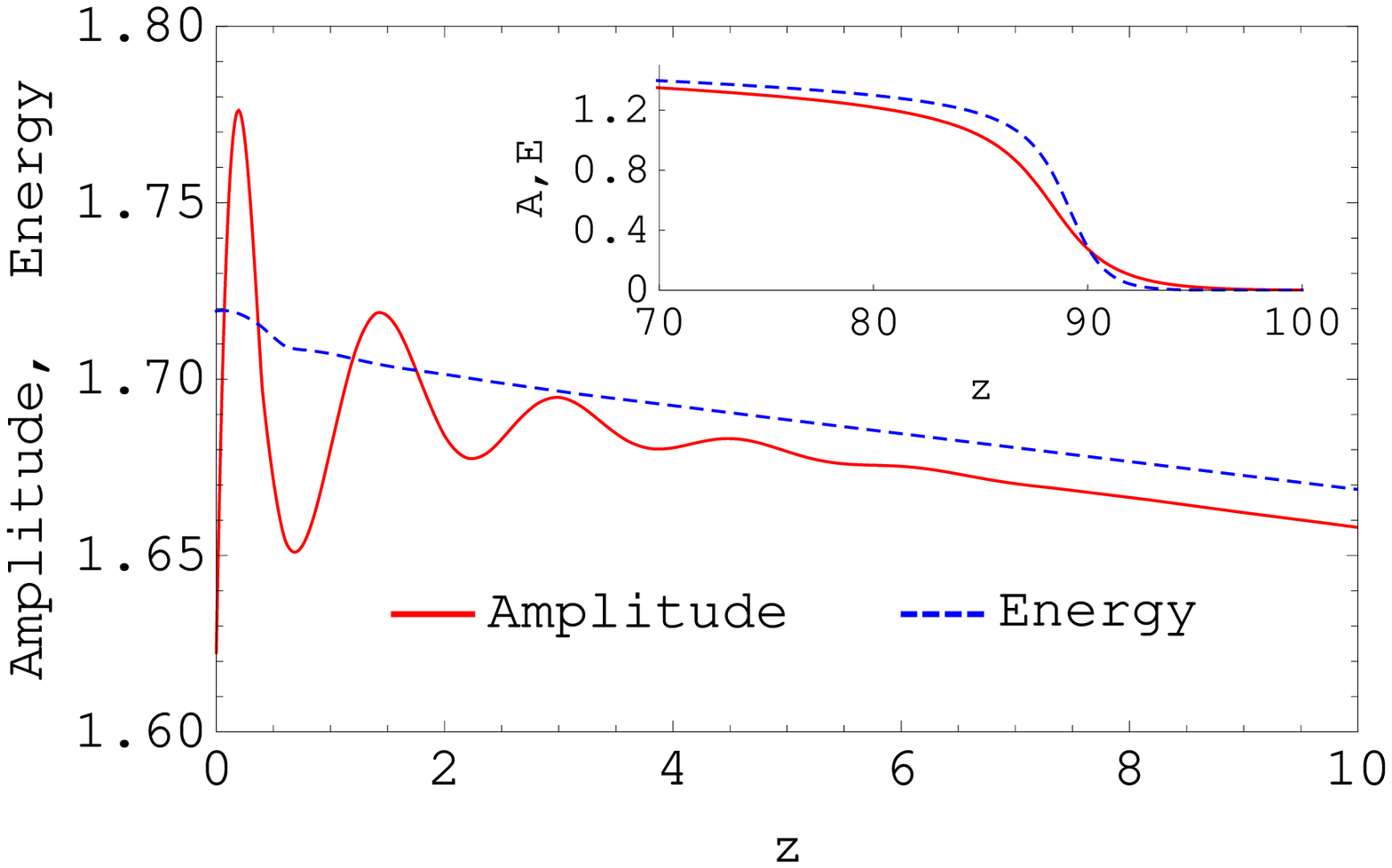}\quad
            \includegraphics[width=5cm,height=6cm,clip]{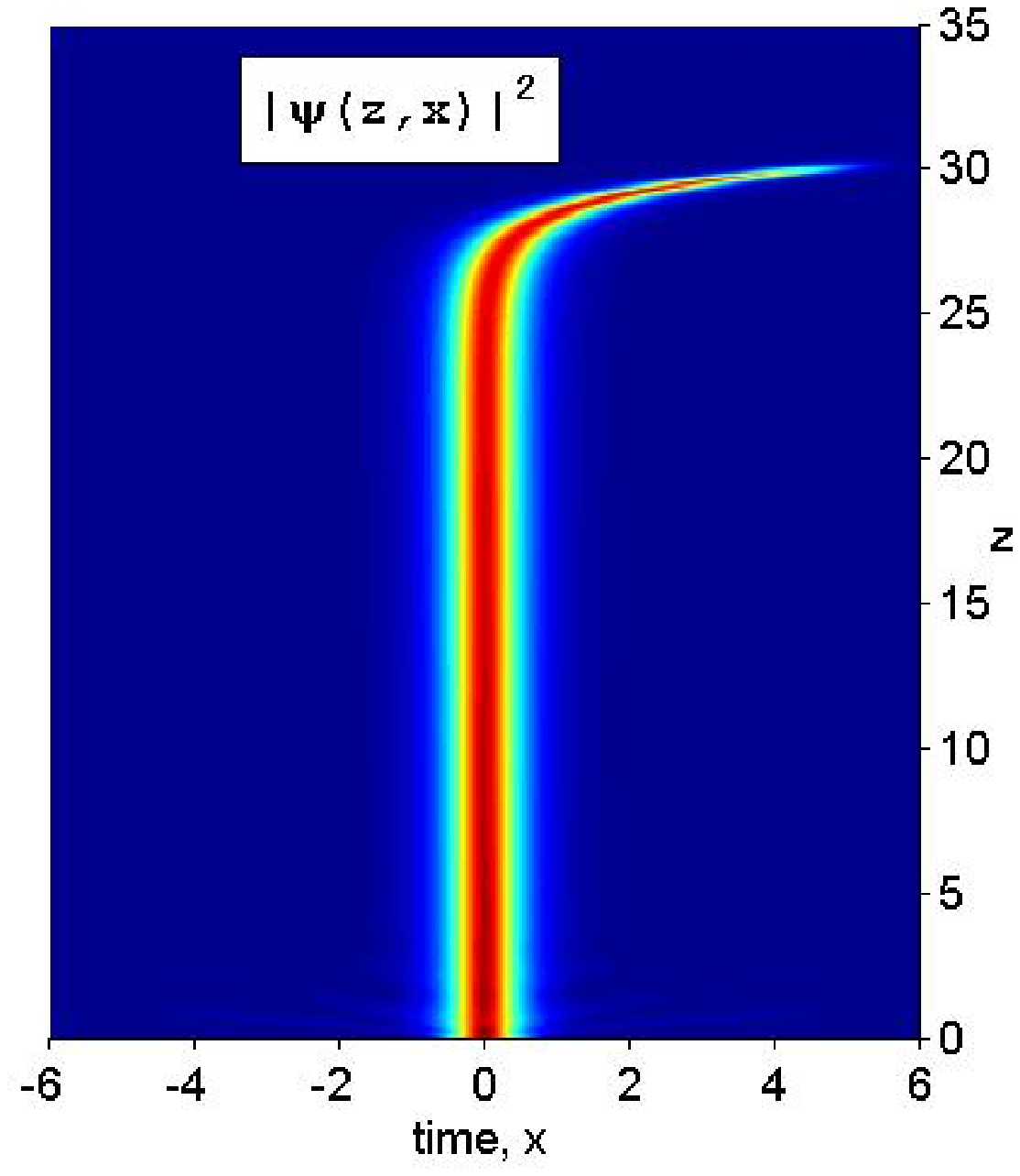} \quad
            \includegraphics[width=5cm,height=6cm,clip]{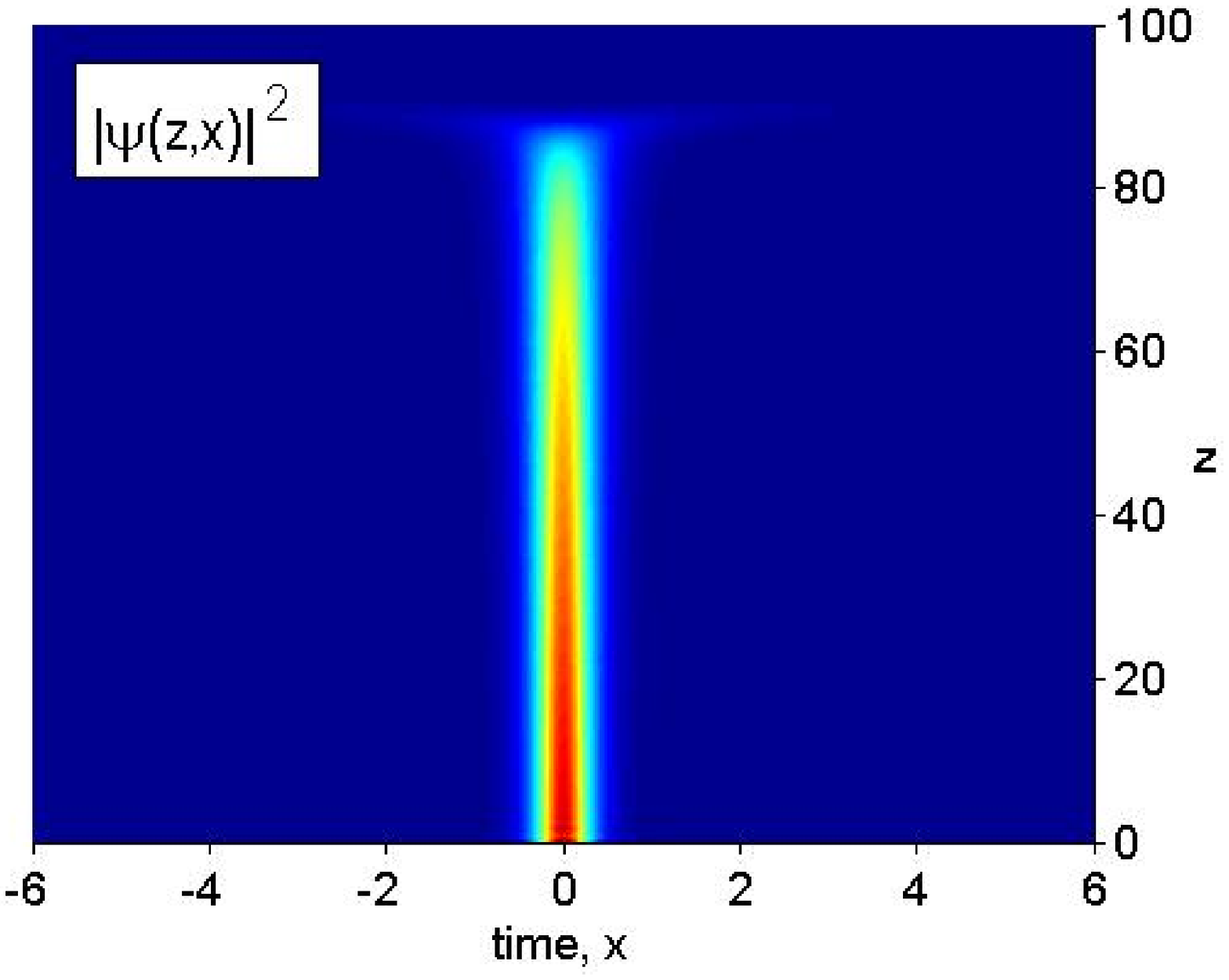}}
\caption{Left panel: The amplitude and energy of the soliton decay
as it propagates along the fiber due to the wave tunneling effect.
The inset shows the nonlinear character of the energy loss at
longer distances and abrupt disintegration of the pulse at some
critical energy. The data is obtained by numerical simulation of
Eq. (\ref{averaged}) with absorbing boundary conditions. The
initial wave is a Gaussian pulse with the amplitude $A=1.6227$ and
width $\tau=0.521$, as predicted by VA. Middle panel: The soliton
slightly displaced from the top of the inverted parabola (by
amount $\Delta x \sim \pm 10^{-3}$) slides down with increasing
velocity in the corresponding direction. This instability develops
even due to a numerical noise, when the soliton is placed exactly
at the origin $x=0$. Right panel: When calculation is performed
with centering of the pulse, a complete disintegration can be
observed at sufficiently long propagation distance.} \label{fig6}
\end{figure}

To verify the above conclusions following from our model, we
performed numerical simulations of the pulse propagation governed
by Eq. (\ref{averaged}). When we introduce the wave profile
predicted by VA, which slightly differs from the solution of Eq.
(\ref{ivp}), as initial condition to Eq. (\ref{averaged}), the
pulse quickly adjusts itself by performing damped oscillations of
its amplitude, and then continuously decay both in the energy and
amplitude, as shown in Fig. \ref{fig6}. The frequency of
oscillations of the amplitude well agrees with the prediction of
VA, when we expand the potential (\ref{pot1}) near its minimum
$\tau_0$ and estimate the corresponding frequency and period
$\omega_0=\sqrt{d^2 U/d \tau^2|_{\tau=\tau_0}} \simeq 5.12$,
$T_0=2\pi/\omega_0=1.22$. However, the VA does not take into
account the dissipative effects. In numerical simulations we use
the absorbing boundary technique \cite{berg} to prevent the
interference of the soliton with the linear waves, otherwise
reflected from the integration domain boundaries. To calculate the
energy outflow from the soliton, we monitor the amount of energy
in the central part of the domain ($x \in [-2,2]$ in Fig.
\ref{fig6}) where the bulk of the pulse is confined. The energy
loss rate $dE/dz$ is characteristic for the fiber parameters and
initial pulse power, which define the coefficients of the averaged
NLSE (\ref{averaged}). The instability of the center-of-mass
position of the soliton is demonstrated in the middle panel of
Fig.~\ref{fig6}. If the calculation is performed with centering of
the soliton, when at each step its center-of-mass is held at the
origin, we observe complete disintegration of the pulse due to the
energy loss, as shown in the right panel of this figure. The
critical energy, at which the pulse disintegrates, according to
the inset of the left panel is $E_{cr} \simeq 1.2$. This numerical
finding is in good agreement with the prediction of VA Eq.
(\ref{ecrit}) $E_{cr} = 1.24$.

\begin{figure}[htb]
\centerline{\includegraphics[width=6cm,height=6cm,clip]{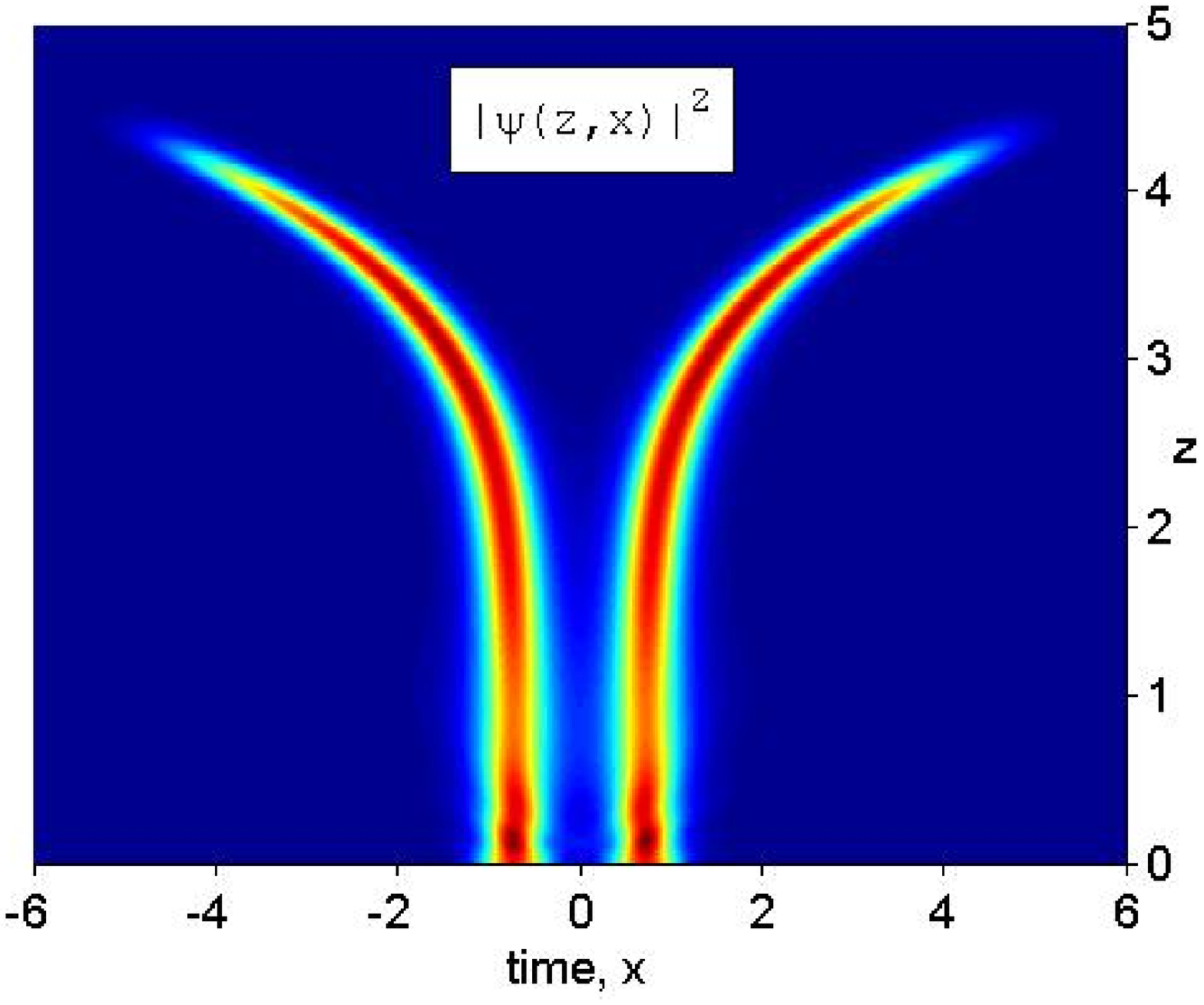}\quad
            \includegraphics[width=6cm,height=6cm,clip]{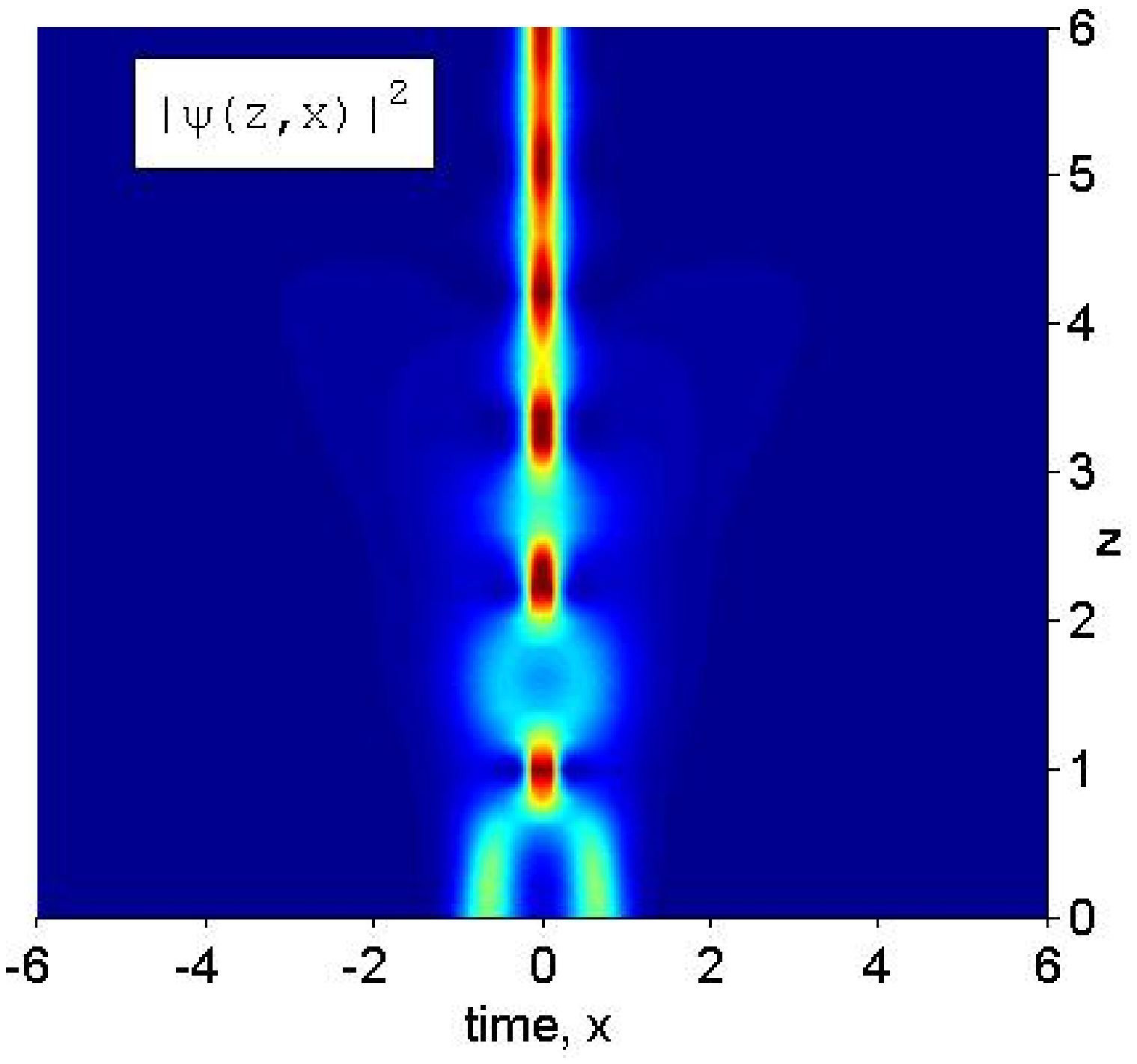} \quad
            \includegraphics[width=6cm,height=6cm,clip]{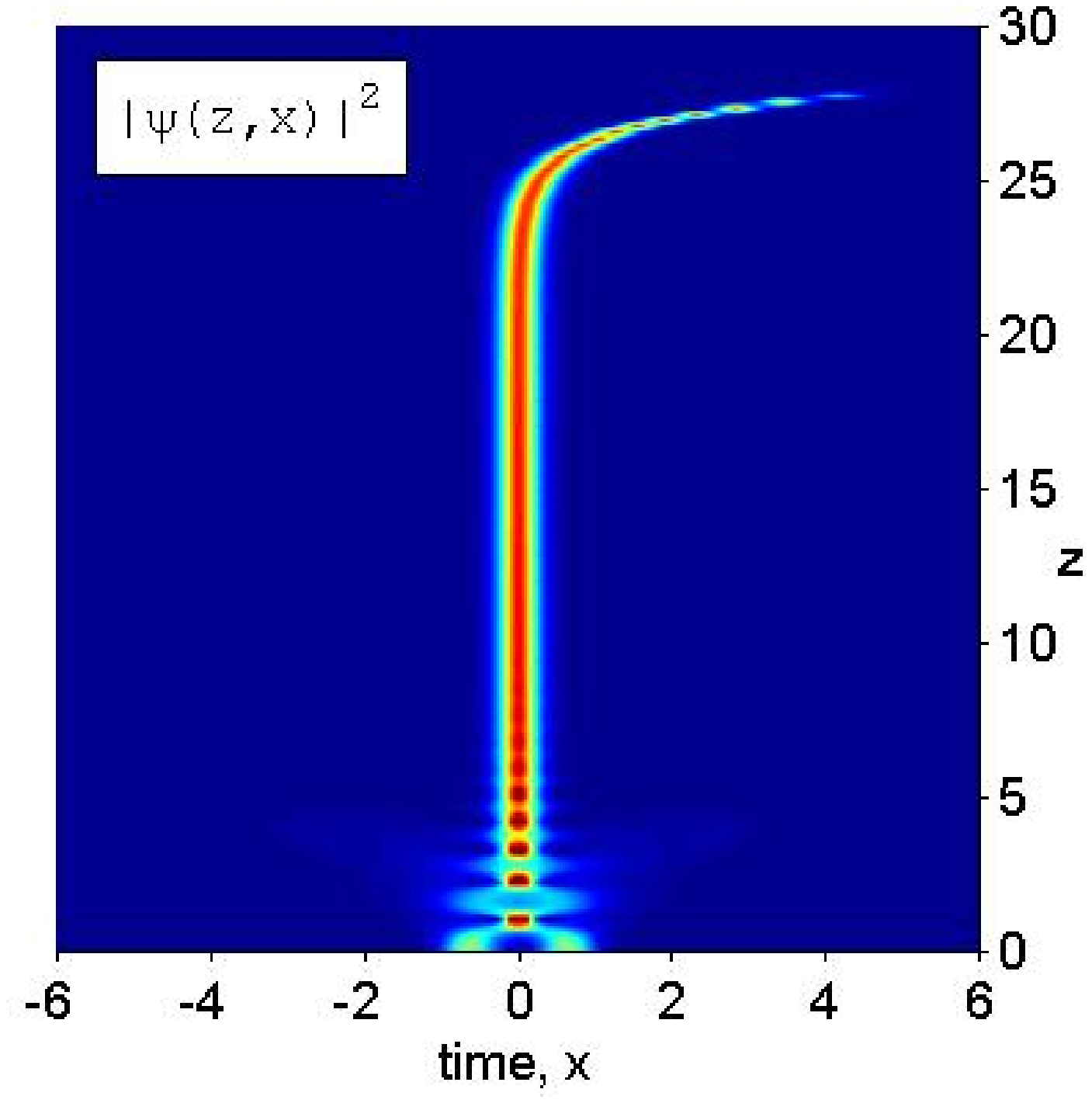}}
\caption{Left panel: Two in-phase solitons placed sufficiently far
from each-other ($\Delta x=0.72$) move apart. Middle panel: When
solitons are placed at a distance ($\Delta x=0.65$) which is less
than some critical value, they collide few times and merge. Right
panel: Long distance behavior of the coalescent pulse (of the
middle panel) is similar to the single soliton case, as its
center-of-mass develops instability and slides down the inverted
parabola.} \label{fig7}
\end{figure}

In fact the above mentioned mechanism of energy loss sets the
limit to attainable robustness of DM solitons and molecules in the
given setup. The instability of the center-of-mass of the soliton
will make it difficult to held the pulse in the middle of its time
slot.
In Fig. \ref{fig7} we demonstrate the propagation of two solitons
governed by Eq. (\ref{averaged}). When two in-phase solitons are
initially  placed at a separation exceeding some critical value,
they move apart with increasing velocity under the effect of the
expulsive parabolic potential. If the solitons are placed at a
smaller temporal distance, they collide few times and merge
together, and later the combined pulse develops instability,
analogous to the single soliton case and slides down the inverted
parabola. Similar behavior of two-soliton states were reported in
the dissipative counterpart of Eq. (\ref{averaged}) in Ref.
\cite{grimshaw2007}. An additional fact relevant to coupled
soliton propagation in this system is that, the out-of-phase
solitons always repel each-other and diverge from the origin in
accelerated manner.

The main result of our study of pulse propagation governed by the
averaged NLSE (\ref{averaged}) is that, it does not support truly
stable solitons and molecules as the original DM Eq. (\ref{nlse3})
does. Rather it shows, that solitons and molecules in the
conservative DM fiber are meta-stable, both in terms of energy and
temporal position. In addition, the model of averaged NLSE allows
to find the existence regimes and identify the limits of stability
of solitons and molecules in DM fibers. The inclusion of
dissipation and gain into the model may change the results to some
degree. This will be a subject for separate study.

\section{Conclusions}

We have developed a variational approximation which successfully
describes the propagation of soliton molecules in DM fibers. The
pulse shapes for a two-soliton molecule and three-soliton
molecule, predicted by VA are shown to be sufficiently close to
the numerically exact shapes found by solution of the original DM
NLS equation. Then we have studied the dynamics of solitons and
molecules in the averaged NLSE corresponding to the selected DM
fiber system. The approach of averaged NLSE allows to identify the
regimes of existence of solitons and molecules in the original DM
system, and reveal the fundamental source of instability of
soliton propagation in DM fibers, which is linked to continuous
outflow of energy from the pulse due to the wave tunneling
phenomenon. The model also predicts the instability of the
temporal position of the pulse within its time slot. All
calculations are performed using the parameters of the existing DM
fiber setup \cite{rohrmann2012,rohrmann2013}. The model may
provide guidance in further studies of the properties of soliton
molecules in DM fibers.

\section*{Acknowledgments}
We thank F. Mitschke, A. Hause and E. N. Tsoy for valuable
discussions. U.A.K. and B.B.B. are grateful to the Department of
Physics of the KFUPM for the hospitality during their stay. This
work is supported by the research grant UAEU-NRF 2011 and KFUPM
research projects RG1214-1~and~RG1214-2.


\begin{thebibliography}{99}

\bibitem{mitschke-book}
Fedor Mitschke, {\it Fiber Optics. Physics and Technology}
(Springer-Verlag, Berlin, Heidelberg, 2009).

\bibitem{stratmann2005}
M. Stratmann, T. Pagel, and F. Mitschke, Phys. Rev. Lett. {\bf
95}, 143902 (2005).

\bibitem{hasegawa1973}
A. Hasegawa and F. Tappert, Appl. Phys. Lett. {\bf 23}, 142
(1973).

\bibitem{mollenauer1980}
L. F. Mollenauer, R. H. Stolen, and J. P. Gordon, Phys. Rev. Lett.
{\bf 45}, 1095 (1980).

\bibitem{hasegawa1995}
A. Hasegawa and Y. Kodama, {\it Solitons in optical
communications} (Clarendon Press, Oxford, 1995).

\bibitem{mollenauer2006}
L. F. Mollenauer and J. P. Gordon, {\it Solitons in Optical
Fibers: Fundamentals and Applications} (Academic Press, San Diego,
2006).

\bibitem{turitsyn2012}
S. K. Turitsyn, G. B. Brandon Bale, and M. P. Fedoruk, Phys. Rep.
{\bf 521} 135 (2012).

\bibitem{hause2008}
A. Hause, H. Hartwig, M. B\"ohm, and F. Mitschke, Phys. Rev. A
{\bf 78}, 063817 (2008).

\bibitem{rohrmann2012}
P. Rohrmann, A. Hause, and F. Mitschke, Sci. Rep. {\bf 2}, 866
(2012).

\bibitem{rohrmann2013}
P. Rohrmann, A. Hause, and F. Mitschke, Phys. Rev. A {\bf 87},
043834 (2013).

\bibitem{maruta2002}
A. Maruta, T. Inoue, Y. Nonaka, and Y. Yoshika, IEEE J. Selected
Topics Quant. El., {\bf 8}, 640 (2002).

\bibitem{anderson1983}
D. Anderson, Phys. Rev. A {\bf 27}, 3135 (1983).

\bibitem{turitsyn1998}
S. K. Turitsyn, I. Gabitov, E. W. Laedke, V. K. Mezentsev, S. L.
Musher, E. G. Shapiro, T. Schafer, and K. H. Spatschek, Opt.
Commun. {\bf 151}, 117 (1998).

\bibitem{pare1999}
C. Pare and P. -A. Belanger, Opt. Commun. {\bf 168}, 103 (1999).

\bibitem{feng2004}
B. -F. Feng and B. A. Malomed, Opt. Commun. {\bf 229}, 173 (2004).

\bibitem{cautaerts2000}
V. Cautaerts, A. Maruta, and Y. Kodama, Chaos {\bf 10}, 515
(2000).

\bibitem{malomed-book}
B. A. Malomed, {\it Soliton management in periodic systems}
(Springer, 2006).

\bibitem{turitsyn1997}
S. K. Turitsyn, JETP Letters {\bf 65} 845 (1997); S. K. Turitsyn,
T. Sch\"afer, K. H. Spatschek, V. K. Mezentsev, Opt. Commun. {\bf
163}, 122 (1999).

\bibitem{nijhof2000}
J. H. B. Nijhof, W. Forysiak, and N. J. Doran, IEEE J. Select.
Topics Quant. El., {\bf 6}, 330 (2000).

\bibitem{kodama1997}
Y. Kodama, S. Kumar, and A. Maruta, Opt. Lett. {\bf 22}, 1689
(1997).

\bibitem{spatschek1995}
K. H. Spatschek, S. K. Turitsyn and Y. S. Kivshar, Phys. Lett. A
{\bf 204}, 269 (1995).

\bibitem{hasse1982}
R. Hasse, Phys. Rev. A {\bf 25}, 583 (1982).

\bibitem{moores1996}
J. D. Moores, Opt. Lett., {\bf 21}, 555 (1996).

\bibitem{nakkeeran2001}
 K. Nakkeeran, J. Phys. A: Math. Gen. {\bf 34}, 5111 (2001).

\bibitem{mak2005}
 C. C. Mak, K. W. Chow and K. Nakkeeran, J. Phys. Soc. Japan {\bf 74}, 1449
 (2005).

\bibitem{xu2003}
Zhiyong Xu, Lu Li, Zhonghao Li, Guosheng Zhou, and K. Nakkeeran,
Phys. Rev. E {\bf 68}, 046605 (2003).

\bibitem{grimshaw2007}
R. Grimshaw, K. Nakkeeran, C. K. Poon, and K. W. Chow, Phys. Scr.
{\bf 75}, 620 (2007).

\bibitem{carr2002}
L. D. Carr and Y. Castin, Phys. Rev. A {\bf 66}, 063602 (2002).

\bibitem{berg}
P. Berg, F. If, P. L. Christiansen, and O. Skovgaard, Phys. Rev. A
{\bf 35}, 4167 (1987).

\end{thebibliography}
\end{document}